\documentclass[12pt]{article}
\usepackage[square,authoryear]{natbib}
\usepackage{marsden_article}
\usepackage{epstopdf}

\newcommand{\ns}{\mspace{-1.5mu}}             
\newcommand{\ps}{\mspace{1.5mu}}             

\begin{document}

\title
{\Large \bf Concatenating Variational Principles and the Kinetic Stress-Energy-Momentum Tensor
\\[1.5ex]}
\author{%
{\bf Marco Castrill\'on L\'opez} \\
Departamento de Geometr\' ia y Topolog\' ia\\[-2pt]
Facultad de Ciencias Matem\' aticas \\[-2pt]
Universidad Complutense de Madrid \\[-2pt]
28040 Madrid, Spain
\\
\and 
{\bf Mark J. Gotay} \\
Department of Mathematics\\[-2pt]
University of Hawai`i\\[-2pt]
Honolulu, Hawai`i 96822, USA \\
\and
{\bf Jerrold~E.~Marsden} \\
Control and Dynamical Systems 107-81\\[-2pt] 
California Institute of Technology\\[-2pt]
Pasadena, California 91125, USA\\[12pt]
}

\date{26 November 2007}

\maketitle

\begin{abstract}
We show how to ``concatenate'' variational principles over different bases into one over a single base, thereby providing a unified Lagrangian treatment of interacting systems. As an example we study a Klein--Gordon field interacting with a mesically charged particle. We employ our method to give a novel group-theoretic derivation of the kinetic stress-energy-momentum  tensor density corresponding to the particle.  
\end{abstract}

\newpage


\section{Introduction and Setup}

Let us recall the geometric setting of a classical variational principle (\cite{GIMMsyI}): We are given a  fibration $Y \to X$, with $\dim X = n+1$, and we wish to extremize an action of the form 
\begin{equation*}
S(\psi) = \int_X \mathcal{L} (j^1\psi)
\end{equation*}
where $\psi: X \to Y$ is a section and $\mathcal{L} : J^1Y \to \Lambda^{n+1}X$ is a specified Lagrangian density.\footnote{\, For simplicity we consider only first order theories. We also ignore technical issues and proceed formally.}

One commonly encounters several (say $K$) such variational principles simultaneously, for instance when one studies the Newtonian dynamics of a swarm of charged particles (in a background electromagnetic field), or the interaction between Dirac and Yang--Mills  fields. In the cases cited, the relevant fibrations have the form $Y_i \to X$ for the $i^{\textup{th}}$ variational principle; the key point being that each fibration has the \emph{same} base $X$. To combine these variational principles into a single principle is a straightforward matter: one builds the fiber product $Y_1 \times_X \cdots \times_X Y_K \to X$, and then on the first jet of this bundle one takes as the Lagrangian density an expression of the form $\mathcal{L} _1 + \cdots + \mathcal{L} _K + \mathcal{L} _{\textup{int}}$ for some interaction terms $\mathcal{L} _{\textup{int}}$.

It is less clear how to deal with variational principles with disparate bases, that is, fibrations $Y_i \to X_i$ in which the $X_i$ are all different. A simple example is a nucleon  moving in a dynamic Klein--Gordon field. (Here the configuration bundle for the nucleon is $X \times \mathbb{R}  \to \mathbb{R} $, where $X$ is 4-dimensional  spacetime and  $\mathbb{R} $ is the material world line of the nucleon. The fibration for the Klein--Gordon field is $\mathbb{R}  \times X \to X$, sections of which are scalar fields on spacetime.) Even if the bases are identical, it may be desirable to distinguish them. This is the case, for instance, in relativistic multiparticle systems (cf. \cite{Anderson1967}), when one wants to parametrize each particle's trajectory by its own proper time, as opposed to a single ``universal'' time.

In this context of disparate bases, one standard way to proceed is as follows. Construct an action functional using sections $\psi _i : X _i \rightarrow Y _i$ for the $i^{\rm th}$ bundle by setting
\begin{equation} \label{GeneralAction}
S ( \psi _i,  \ldots , \psi _K ) = \sum _{i = 1 }^{K } \int _{X _i } \mathcal{L} _i (j^1\psi_i) + \int _{X_1 \times \cdots \times  X_K} \mathcal{L} _{\textup{int}} (j^1\psi_1,\ldots,j^1\psi_K).
\end{equation}
Then varying these fields $\psi _i$, one obtains the Euler--Lagrange equations for the problem. See equation \eqref{MesonAction} for a specific example.

However, while producing the Euler--Lagrange equations, this approach has the unsatisfactory feature of not yielding a field theory in the usual sense, in which the fields are sections of a single bundle and which has a well-defined Lagrangian density. This or some other formalism is needed if one wishes to tap into the machinery of multisymplectic geometry, multimomentum maps, stress-energy-momentum (``SEM'') tensors, and constraint theory, etc. 
\medskip

To \emph{concatenate} variational principles with disparate bases in such a way as to recapture a genuine field theory, we proceed as follows. To begin, construct the product bundle $Y_1 \times \cdots \times Y_K \to X_1 \times \cdots \times  X_K$, which we denote $Y\to X$ for short. In agreement with experience we restrict attention to \emph{product sections} of this bundle of the form $\psi = (\psi_1,\ldots , \psi_K)$, where each $\psi_i$ is a section of $Y_i \to X_i$. 
With $\psi = (\psi_1,\ldots , \psi_K)$ such a  section,
$$j^1\psi(x) = \big(j^1\psi_1(x_1),\ldots,j^1\psi_K(x_K) \big)$$ 
where $x = (x_1,\ldots,x_K)$. Denote by $\bar J^1Y$ the subbundle of $J^1Y$ consisting of all such jets; equivalently, $\bar J^1Y =  J^1Y_1 \times \cdots \times J^1Y_K$.

Given Lagrangian densities $\mathcal{L} _i$ on the jet bundles  $J^1Y_i$, it is simple enough to lift them to maps, still denoted by $\mathcal {L}_i$, on the concatenated jet bundle $\bar J^1Y$ by composing with projections:
$$j^1\psi(x) \mapsto \mathcal {L}_i \big(j^1\psi_i(x_i) \big).
$$
	But how do we  concatenate these ${\mathcal{L} }_i$ into a \emph{single} Lagrangian density? Even ignoring interaction terms, we cannot just add the ${\mathcal{L} }_i$ as they take values in different spaces, viz. $\Lambda^{n_i+1}X_i$ and so need not be forms of equal rank. The trick is to ``suspend'' the ${\mathcal{L} }_i: \bar J^1Y \to \Lambda^{n_i+1}X_i$ to maps $\bar J^1Y \to \Lambda^{N+K}X$, where $N = n_1+ \cdots +n_K$, by  inserting suitable tensor densities in the ${\mathcal{L} }_i$ to ``even out'' their ranks in the target. 

First, we pull $ \mathcal {L}_i $ back via the projection $X \to X_i$ to an $(n_i+1)$-form on $X$.  Second, for each $i$ choose scalar densities $\mathfrak D_i$ of weight 1 on $X_1 \times \cdots \widehat{X_i} \cdots \times X_K$. Now in $ \mathcal {L}_i  = L_i d^{\ps n_i +1}\ns x_i$ the coefficient $L_i$ transforms as a scalar density of weight 1 on $X_i$, so the coefficient in
\begin{align*}
\mathcal {\bar L}_i : & \!\! =   L_i \ps d^{\ps n_i + 1}\ns x_i  \wedge  \mathfrak D_i \ps
d^{\ps n_1 +1}\ns x_1   \wedge    \cdots \wedge \widehat{d^{\ps n_i +1}\ns x_i }\wedge \cdots \wedge d^{\ps n_K+ 1}\ns x_K \\[2ex]
& \!\!  =   \pm L_i \ps  \mathfrak D_i
 \ps d^{\ps N+K}\ns x 
\end{align*}
will also transform as a scalar density of weight 1 on $X$ under the subgroup 
$$\textup{Diff}(X_1) \times \cdots \times \textup{Diff}(X_K) \subset \textup{Diff}(X)
$$ 
(which is sufficient for our purposes). The densities $\mathfrak D_i$ are to be chosen by hand, depending on the precise structure  of the system; see the examples  in \S\S  \ref{MesonSection} and \ref{ExamplesSection}.
Thus modified, we may assemble $\bar{\mathcal{L} }_1 + \cdots + \bar{\mathcal{L} }_K$ into a map 
\[
\bar{\mathcal{L} } : \bar J^1Y \to \Lambda^{N+K}X.
\]
 Interaction terms, which are typically defined over several of the bases $X_i$ (again, see the following examples) are treated similarly. Finally, it is straightforward to deal with composite situations in which some of the bases are identical and others are not.

Ultimately, the specific choice of the $\mathfrak D_i$ will not matter as long as
$$\int_{X} L_i \ps \mathfrak D_i \,d^{\ps N+K}\ns x
= \int_{X_i} L_i \,d^{\ps n_i+1}\ns x$$ for each $i$, that is, the concatenated action reduces to the original action. Specifically, this means that
\[
\int _X \bar{\mathcal{L} }(j^1\psi) = S ( \psi ) 
\]
where the right hand side is given by \eqref{GeneralAction}. \emph{In particular, the Euler--Lagrange equations remain unaltered when the Lagrangian $\bar{\mathcal{L} }$ is used in place of the action functional \textup{\eqref{GeneralAction}.}} 

Once we have a total Lagrangian density in hand (albeit possibly a distributional one), we may proceed in the usual fashion. Thus we may compute the equations of motion and various geometric objects, such as SEM tensors. To extract  physical information from these objects, however,  it will normally be necessary to ``project'' them from $X$ to some $X_i$ or products thereof; this projection is accomplished by integration over the remaining $X_j$. 

Rather than continuing to try to describe the procedure in generality, it is more instructive to illustrate it via a simple example. (It really is easier done than said!)

\medskip

In \S\ref{MesonSection} we apply this method to a system consisting of a Klein--Gordon field interacting with a mesically charged particle. (Think of a pion field  interacting with a nucleon.) Beyond illustrating concatenation, this example has interesting features which are worth elucidating. In particular, we study  the SEM  tensor density of this system. Its  computation, following \cite{GoMa1992}, is interesting in that it naturally produces the Minkowski, or kinetic, SEM tensor for a moving particle as a matter of course. To our knowledge, this SEM tensor has never been derived via a Lagrangian from first principles;  it has always been inserted into the formalism in an \emph{ad hoc} manner. An important point therefore is that our method is not merely a `tidy' means of packaging variational principles; it is capable of providing, in an entirely straightforward fashion, quantities which otherwise cannot be obtained except in makeshift  ways.

Finally in \S\ref{ExamplesSection} we briefly indicate some other contexts in which our results should be useful.

\section{Motion of a Mesically Charged Particle in a Klein--Gordon Field} \label{MesonSection}

 Let $X$ be an oriented spacetime with metric $G$. We consider a real Klein--Gordon field $\phi: X \to \mathbb{R} $ of mass $M$ interacting with a particle of mass $m$ and mesic charge $\varepsilon$. The particle's trajectory in spacetime (or ``placement field'') is $z: \mathbb{R}  \to X$. The base for the system is thus $X \times \mathbb{R} $, the second factor being thought of as a time axis,\footnote{\, Not necessarily proper time.} and the configuration bundle $Y$ is then 
$$ (\mathbb{R}  \times X) \times (X \times \mathbb{R} )  \to X \times \mathbb{R} $$ 
with coordinates $(\phi, X^a )$ on the fiber and $ (x^\mu,\lambda) $ on the base. We set $z^a = X^a \circ z$. 

Our presentation is based upon the excellent exposition in Chapter 8 of \cite{Anderson1967}, to which we refer the reader for further information. The action \eqref{GeneralAction} for the system in this case is usually written
\begin{align}
 S(\phi,z) & =   \int_X \frac12 \Big(G^{\mu\nu}(x) \phi_{,\mu}(x) \phi_{,\nu}(x) - M^2\phi(x)^2 \Big) \sqrt{-G(x)}\,d^{\ps 4}\ns x  \nonumber \\[2ex]
 &  \quad  - \;  \int_{X\times \mathbb{R} } \varepsilon \phi(x) \|\dot z(\lambda)\| \, \delta^4(x - z(\lambda)) \,d^{\ps 4}\ns x \, d\lambda \nonumber \\[2ex]
&  \quad - \; \int_{\mathbb{R} } m  \|\dot z (\lambda)\| \, d\lambda,
\label{MesonAction}
\end{align}
where the dot denotes differentiation with respect to  $\lambda$ and $\|\dot z\| = \sqrt{-G_{ab}\dot z^a \dot z^b}$. Observe that the bases for the free Klein--Gordon term and the free particle term are different, and that the interaction term in the middle lives on the product of these. 
\medskip 

Before proceeding, there are two technical issues that need to be resolved, stemming from the presence of the two factors of $X$ in the configuration bundle. First,  note that in the leading term of $S$, $G$ is regarded as living on the $X$ in the base, while in the last term it evidently resides on the $X$ in the fiber. It is necessary to know precisely where $G$ lives, as this has an effect on the subsequent analysis: if on the base, then $G$ is treated as a field, while if on the fiber it is simply thought of as a geometric object. We reconcile these two interpretations by taking $G$ to be anchored to the base, and then pulling it back to the fiber by means of the following construction.\footnote{\ This is a variant of the Kucha\v r method of parametrizing a classical field theory; see \cite{GoMa2008} and \cite{CLGoMa2008} for details.}$^,$\footnote{\ At the end of this section we will briefly examine what happens if instead we anchor $G$ to the fiber.} Introduce yet \emph{another} factor of $X$ in the fiber along with diffeomorphisms $\eta: X \to X$, viewed as sections of $X \times X  \to X$, with corresponding configuration and multivelocity variables $\eta^a = X^a \circ \eta$ and $\eta^a{}_\mu = \partial (X^a \circ \eta)/\partial x^\mu$, respectively. (We can, and do, regard the two copies of $X$ in the fiber as identical.) We use these auxiliary nondynamic fields, the \emph{covariance fields}, to (\emph{i}) identify the copies of $X$ in the fiber with that in the base, and (\emph{ii}) endow the new copy of $X$ in the fiber with the metric $g = \eta_*G$ with components
$$g_{ab}  = G_{\mu\nu}\kappa^{\mu}{}_a \kappa^{\nu}{}_b ,$$ 
where $(\kappa^\mu{}_a) = (\eta^a{}_\mu)^{-1}.$ All this is summarized in the figure below. 
\begin{figure}[ht]
\begin{center}
\includegraphics[scale=0.65,angle=0]{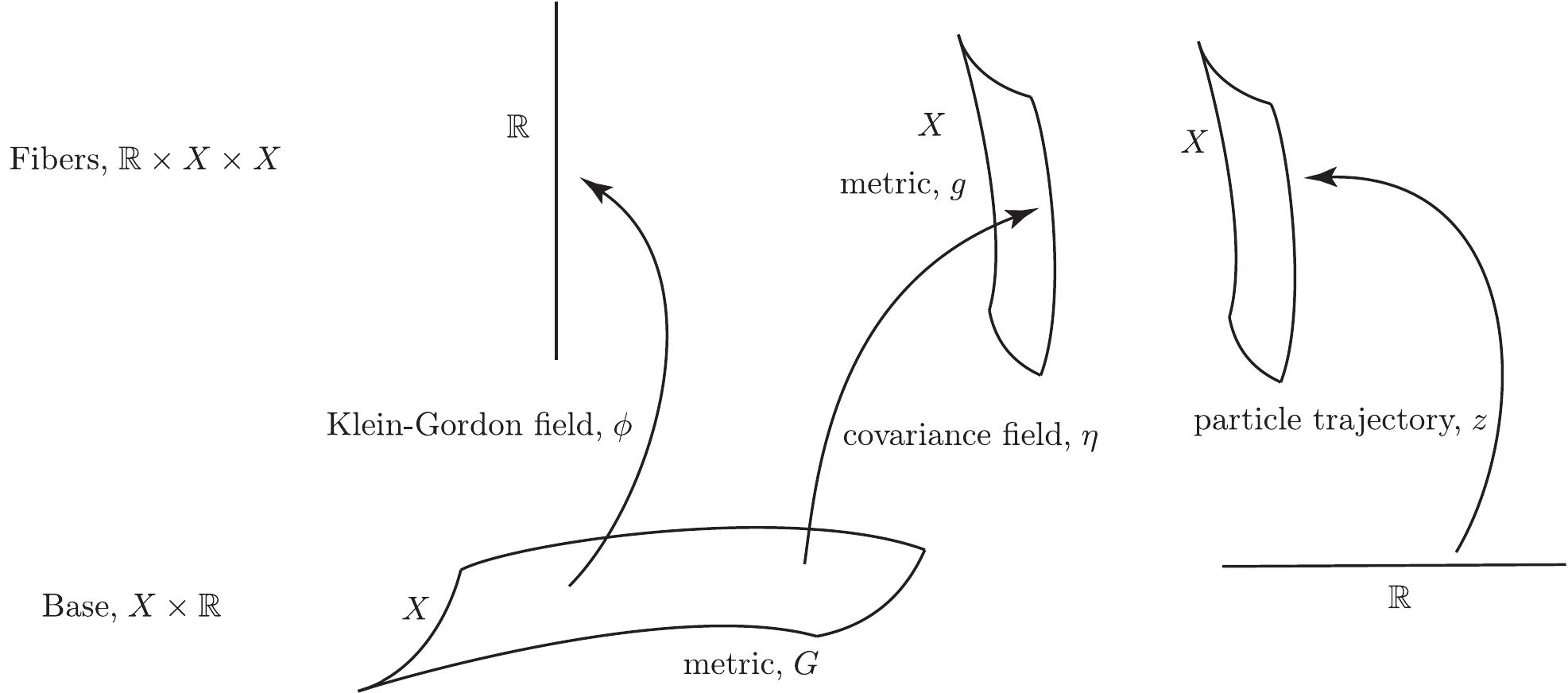}
\end{center}
\begin{center}
{\footnotesize {The general set up for the introduction of covariance fields.}}
\end{center}
\end{figure}

Second,  the delta function $\delta^4(x - z(\lambda))$ must be modified, as it compares elements $x$ in the base with elements $z(\lambda)$ in the fiber. As just indicated we can use the covariance fields to remedy this problem as well: we need only write $\delta^4\big(x- \eta^{-1}(z(\lambda))\big)$ instead. It is sometimes convenient to replace
\begin{equation}
\delta^4 \big(x- \eta^{-1}(z(\lambda))\big) = \delta^4(\eta(x) - z(\lambda))(\det \eta_*) 
\label{df}
\end{equation}
using the properties of delta functions (cf. the Appendix), where $\eta_*$ is the Jacobian of $\eta$. From this we see that $\delta^4 \big(x- \eta^{-1}(z(\lambda))\big)$ (\emph{i})  is a scalar density on $X$ (again, see the Appendix), and ($ii$) depends upon the spacetime derivatives of $\eta$, even though this is not obvious at first glance. The reason we do not insist on a fixed identification of the base $X$ with the fiber $X$, and instead allow a variable identification by means of the covariance fields, will become apparent in due course.

\paragraph{Remark.}
Analogous fields $\eta$, called covariance fields as well, are introduced in \cite{GoMa2008} and \cite{CLGoMa2008}, but there they have a different purpose, namely, to make a field theory on a given background generally covariant and in doing so, they are introduced as \emph{dynamic} fields.
\hfill $\blacklozenge$
\bigskip 

In addition to the covariance fields $\eta$, we introduce a (positive-definite) metric $K$ on $\mathbb{R} $ as a nondynamic field. We suppose that $K$ is chosen so that $\mathbb{R} $ has metric volume 1.
\medskip

With these fixes we may now concatenate the three action terms over the composite base $X \times \mathbb{R} $ as $ S(\phi,z) = \int_{X\times \mathbb{R} } \bar L \, d^{\ps 4}\ns x \, d \lambda$, with the Lagrangian 
\begin{eqnarray}
\bar L(x^\mu,\lambda, \phi,\phi_{,\mu},z^b,\dot z^b\hspace{-2ex} & \hspace{-2ex} ; \hspace{-2ex}  & \hspace{-2ex}  \eta^a,\eta^a{}_\mu,G_{\rho\sigma}, K) \nonumber \\[2ex] 
&  =  & \frac12  \Big(G^{\mu\nu} \phi_{,\mu} \phi_{,\nu} - M^2\phi^2 \Big)  \sqrt{-G}\,\sqrt{K} 
\nonumber \\[2ex] 
&  & - \,
 \big(m + \varepsilon \phi\big) \|\dot z\|\, \delta^4(\eta- z) (\det \eta_*).
 \label{lag}
\end{eqnarray}
Notice that the interaction term itself needs no essential modification, as the corresponding term in \eqref{MesonAction} is already an integral over $X \times \mathbb{R} $,  but informs our choice of 
scalar density  in the free particle term, viz.  $\delta^4 \big(x- \eta^{-1}(z(\lambda))\big)$,  when we suspend the latter to $X \times \mathbb{R} $. 
We have also written this delta function in the form \eqref{df} to make it clear that $\bar L$ is defined pointwise.

\paragraph{Remark.}
The choice of scalar density $\mathfrak D = \sqrt{K}$ in the Klein--Gordon term is hardly unique;  all we require is that $\int_{\mathbb{R} } \mathfrak D\, d\lambda = 1.$ For instance, we could instead take $\delta(\lambda)$ for $\mathfrak D$ with no essential difference.  
\hfill $\blacklozenge$
\bigskip 

As evident from \eqref{lag}, the modified configuration bundle is taken to be 
$$Y' = Y \times_X X^2 \times_X \textup{Lor}(X) \times_{\mathbb{R} } \textup{Riem}(\mathbb{R} ),$$
where we abbreviate the bundle $X \times X  \to X$ by $X^2$, $\textup{Lor}(X)$ is the bundle whose sections are Lorentz metrics on $X$ and, similarly, $\textup{Riem}(\mathbb{R} )$ is the bundle whose sections are Riemannian metrics on $\mathbb{R} $. However, in our approach $\phi$  and $z$ are variational, while $\eta$, $G$ and $K$ are nondynamic fields.  As per the above, we now regard 
\begin{equation*}
\|\dot z\| = \sqrt{ - G_{\mu\nu}\kappa^{\mu}{}_a \kappa^{\nu}{}_b \dot z^a \dot z^b}.
\end{equation*}
\paragraph{Remark.}
Occasionally,  as in \cite{LaLi1979}, one encounters  what one might call ``noncovariant concatenations.'' In the current example, this amounts to writing the terms in the action as integrals over $X$ alone and is effectively accomplished by imposing the coordinate condition $x^0 = \lambda$. 
As this  procedure is not covariant, it can lead to problems (\cite{Leclerc2006a}).
\hfill $\blacklozenge$

\bigskip 


We compute the Euler--Lagrange equations. Varying with respect to  $\phi$ and employing \eqref{df}, we obtain
\begin{align*}
 - M^2 \phi(x) \sqrt{-G(x)}\,\sqrt{K(\lambda)} &   -   \varepsilon \|\dot z(\lambda) \| \,  \delta^4\big(x - \eta^{-1}(z(\lambda))\big) \\[2ex]
&  -   \partial_\mu \! \left( G^{\mu\nu} \phi_{,\nu}  \sqrt{-G} \ps \right)\! (x) \, \sqrt{K(\lambda)}= 0.
\end{align*}
Integrating with respect to  $\lambda$, using the fact that $\textup{vol}_K(\mathbb{R} ) =1$, and rearranging, this reduces to the Klein--Gordon equation
\begin{equation}
\nabla^\mu\nabla_\mu \phi + M^2\phi = - \rho
\label{KGeqn}
\end{equation}
where $\nabla$ denotes the $G$-covariant derivative and 
\begin{equation*}
\rho(x) = \varepsilon (-G)^{- \frac12} \int_{\mathbb{R} } \| \dot z(\lambda)\| \, \delta^4\big(x -  \eta^{-1}(z(\lambda))\big)\, d\lambda
\label{source}
\end{equation*}
is the source density.

Similarly, varying with respect to  $z$ and employing \eqref{df} yield
\begin{align*}
 &  \displaystyle{\frac{\partial}{\partial z^a}}   \Big[ \big(m   +   \varepsilon \phi(x) \big)  \|\dot z(\lambda)\|\, \delta^4\big(x-  \eta^{-1}(z(\lambda))\big) \Big]
 \\[2ex]
& \qquad \quad \ \  +   \mbox{}  \frac{\partial}{\partial \lambda}  \left[\big(m + \varepsilon \phi(x) \big)\, \frac{g_{ab}(z(\lambda)){\dot z}^b(\lambda)}{\| \dot z(\lambda) \|}\, \big(x-  \eta^{-1}(z(\lambda))\big) \right] = 0.
\end{align*}
Carrying out the differentiation  and then integrating over $X$, some manipulations give
\begin{align}
& \frac{d}{d\lambda}  \Bigg[ \big(m   +   \varepsilon \phi\big(\eta^{-1}(z(\lambda))\big) \big) \frac{g_{ab}
(z(\lambda)) {\dot z}^b(\lambda)}{\| \dot z(\lambda) \|} \Bigg]
\nonumber 
 \\[2ex]
& \qquad =   - \varepsilon \kappa^\mu {}_a\phi_{,\mu}\big(\eta^{-1}(z(\lambda))\big) \| \dot z(\lambda) \| \nonumber \\[2ex] 
&  \qquad \qquad+ \mbox{ } \big(m + \varepsilon \phi\big(\eta^{-1}(z(\lambda))\big)\big) \left(
 \displaystyle{ \frac{g_{bc,a}(z(\lambda))\dot z^b(\lambda) \dot z^c(\lambda)}{2\|\dot z(\lambda)\|}} \right). 
 \label{parteqn}
\end{align}
To give insight into these equations, note that in the special case when $(X,G)$ is Minkowski spacetime, $\eta = \textup{Id}_X$, and $\lambda$ is taken to be proper time along the particle's world line,  these equations simplify in a global Lorentz frame  to
\begin{equation*}
\frac{d}{d\lambda}  \Big[\big(m +\varepsilon \phi(z(\lambda)) \big) \dot z_a(\lambda) \Big] =  - \varepsilon \phi_{,a}(z(\lambda)).
\end{equation*}
This is the mesic analogue of the Lorentz force law in electrodynamics.

Neither $K$, the $G_{\mu\nu}$, nor the $\eta^a$ have field equations, since they are not variational. Thus one is free to assign them whatever values one wishes in \eqref{KGeqn} and \eqref{parteqn}. Often, however, one has specific values of $G$ and $\eta$ in mind, e.g., the given spacetime  metric for $G$ and $\textup{Id}_X$ for $\eta$. 

\medskip

Turning now to the SEM tensor, let $\textup{Diff}_c(X) \times \textup{Diff}_c(\mathbb{R} )$ (that is, the group of diffeomorphisms that are the identity outside a compact set) act on the modified configuration bundle $Y'$ according to
\begin{equation*}
 (\sigma \times f)\cdot (x,\lambda,\phi, z;\eta, G,K) =
\big(\sigma(x),f(\lambda),\phi, z; \eta,\sigma_*G,f_*K\big).
\end{equation*}
(We assume that all diffeomorphisms are positively oriented.) The Lagrangian density $\bar{\mathcal{L} } = \bar L \, d^{\ps 4} \ns x \, d \lambda$ is then visibly equivariant with respect to  the induced 
action on $\bar J^1Y'$, that is,\footnote{\ Even though the pointwise action of $\textup{Diff}(X)$ on the fiber of the``covariance bundle'' $X \times X \to X$ is taken to be  trivial, its action on \emph{sections} thereof is not: $\sigma \cdot \eta = \eta \circ \sigma^{-1}$. Thus the identification of the factor of $X$ in the base with that  in the fiber can fluctuate, which is why we allow $\eta$ to be variable in the first place. }
\begin{equation*}
  \bar{ \mathcal{L} }\big((\sigma \times f)\cdot j^1(\phi, z;\eta, G,K\big) =
      (\sigma \times f)_*\bar{\mathcal{L} }\big(j^1(\phi, z; \eta , G,K)\big).
\end{equation*}

We may thus use equation (3.12) in \cite{GoMa1992} to  compute the 5-\emph{dimensional} SEM tensor density 
\begin{equation*}
 \mathcal T = 
\left(
\begin{array}{cc}
 \mathcal T^{\mu}{}_{\nu}   &   \mathcal T^4{}_{\nu}    \\
 \mathcal T^{\mu}{}_{ 4}  &   \mathcal T^{4}{}_{ 4} \end{array}
\right)
\end{equation*}
of the interacting system (where $x^4 = \lambda$).\footnote{\  Using the product metric $G \oplus K$ on $X \times \mathbb{R} $, one could also compute $ \mathcal T$ via the Hilbert formula (4.2) in \cite{GoMa1992}. See also \cite{Leclerc2006a}.}
Integrating over $\lambda$ and raising an index, we project out the \emph{spacetime} SEM tensor density:
\begin{eqnarray}
 \mathfrak T^{\mu\nu} & \!\! = \!\! & \int_{\mathbb{R} } \mathcal T^{\mu\nu} d\lambda \nonumber \\[2ex]
 & \!\! = \!\! & \mathfrak t^{\mu\nu}  + (m + \epsilon \phi)\Theta^{\mu\nu}, 
\label{sptmSEM}
\end{eqnarray}
where 
\begin{equation*}
\mathfrak t^{\mu\nu} = - \frac12 \Big((2G^{\mu\alpha}G^{\nu\beta}  - G^{\mu\nu}G^{\alpha\beta})\phi_{,\alpha}\phi_{,\beta} + G^{\mu\nu}M^2\phi^2\Big)\sqrt{-G}
\label{canSEM}
\end{equation*}
is the canonical SEM tensor density of the (free) Klein--Gordon field and
\begin{equation*}
\Theta^{\mu\nu}(x) = \kappa^\mu {}_a \kappa^\nu {}_b  \int_{\mathbb{R} }\frac{\dot z^a(\lambda) \dot z^b(\lambda)}{\|\dot z(\lambda)\|} \, \delta^{\ps 4}\big(x - \eta^{-1}(z(\lambda))\big)\, d \lambda
\end{equation*}
is the \emph{Minkowski} tensor density.
($m\Theta^{\mu\nu}$ is then the \emph{kinetic} SEM tensor density). 
As well, we compute $\mathcal T^{4}{}_{\nu} = 0 = \mathcal T^{\mu}{}_{ 4}$. 
Finally, we find that when integrated over $X$, $\mathcal T^{4}{}_{4}$ is effectively the Klein--Gordon action:
$$\mathfrak T^4{}_4 = \frac{1}{2}  \left( \int_X \Big(G^{\mu\nu}\phi_{,\mu}\phi_{,\nu} - M^2\phi^2\Big)\sqrt{-G}\, d^{\ps 4}\ns x \right)\! \sqrt{K} .$$

\paragraph{Remark.}
The kinetic SEM tensor density is a familiar object in microscopic continuum mechanics, cf. Chapter 8 of \cite{Anderson1967} and \S33 of \cite{LaLi1979}. \cite{Minkowski1908} originally introduced it in flat-spacetime electrodynamics in order to recover the continuity equation $\mathfrak T^{\mu\nu}{}_{,\nu} = 0$ in view of the fact  that $\mathfrak T_{\rm{EM}}^{\mu\nu}{}_{,\nu} \neq 0$ when currents are present.  In the continuum limit of a noninteracting clutch of particles, $\Theta^{\mu\nu}$ goes over to the SEM tensor density for a perfect fluid as in \S\S 9.1-2 of \cite{Anderson1967}. It is interesting that in this limit, the infinite time integrals in the kinetic SEM tensor density disappear and one is left with a local tensor density.

To our knowledge, ours is the first genuine \emph{derivation} of the Minkowski tensor density from first principles in a variational context, once again illustrating the power of multisymplectic geometry in classical field theory and in particular, the usefulness of having a concatenated theory for which one can make use of concepts such as the SEM tensor.

As we have defined it, the Minkowski tensor density depends upon the covariance fields as well as the particle placement field. However, note that when $\eta = \textup{Id}_X$, $\Theta$ reduces to the more familiar expression
\begin{equation*}
\Theta^{ab}(x) =  \int_{\mathbb{R} }\frac{\dot z^a(\lambda) \dot z^b(\lambda)}{\|\dot z(\lambda)\|} \, \delta^{\ps 4}(x - z(\lambda))\, d \lambda.
\end{equation*}
\vskip -32pt
\hfill $\blacklozenge$

\paragraph{Remark.}
Suppose we focus solely on the particle dynamics so that the (original) configuration bundle is $X \times \mathbb{R}  \to \mathbb{R} $. The  corresponding Lagrangian density $- m\|\dot z(\lambda)\| \ps d\lambda$ is $\textup{Diff}_c(\mathbb{R} )$-covariant, and so we may compute the corresponding  SEM \emph{scalar} density as in Example {\bf a},  Interlude II of \cite{GIMmsyII}. We obtain $\mathfrak T = -E$, the energy of the particle, which vanishes as the Lagrangian is time reparametrization-invariant.  (This is reflected by the vanishing of $\mathcal T^4{}_4$ in the 5-dimensional context when there is no Klein--Gordon field.) Thus only when the spacetime $X$ is part of the base of the variational principle do we encounter the kinetic SEM tensor density; it does not appear in standard particle dynamics per se.

To reiterate, even in the absence of other fields, our technique yields yet another (5-dimensional!) treatment of the relativistic free particle that has the advantage of automatically incorporating the Minkowski tensor.
\hfill $\blacklozenge$

\paragraph{Remark.}
Note that the term $\varepsilon \phi\ps \Theta^{\mu\nu}$ in \eqref{sptmSEM}
arises from the interaction of $\phi$ with the mesically charged particle. This term has no analogue in the electrodynamics of particles; there we get simply
$$ \mathfrak T^{\mu\nu}  = \mathfrak T_{\textup{EM}}^{\mu\nu}  + m\Theta^{\mu\nu}.$$ 
Charged strings behave similarly, as we show in \S3A (cf. equation \eqref{semstr}). This can be traced to the fact that the electromagnetic field is a covector, while the Klein--Gordon field is a scalar.
\hfill $\blacklozenge$
\bigskip 

The SEM tensor density $\mathfrak T$ is symmetric. It is also divergence-free, as can be seen from general principles (cf. Proposition 5 in \cite{GoMa1992}). 
One may verify this directly, via a long calculation.

\medskip
We end with a discussion of an alternate treatment of this system. 

Suppose we consider the physical metric as a geometric object $g$ on the fiber as opposed to a field on spacetime. Then we would define $G = \eta^*g$ with components $G_{\mu\nu} = \eta^a{}_\mu \eta^b{}_\nu g_{ab}$. Proceeding as in the above, the Lagrangian density would be
\begin{align*}
\label{newlag}
& \bar L(x^\mu,\lambda, \phi,\phi_{,\mu},z^b,\dot z^b  ;  \eta^a,\eta^a{}_\mu,K) \\[2ex] 
&  \qquad \qquad =   \frac12  \Big(\kappa^\mu {}_a \kappa^\nu {}_b g^{ab} \phi_{,\mu} \phi_{,\nu} - M^2\phi^2 \Big) \sqrt{-g} \, (\det \eta_*)\,\sqrt{K}  
\nonumber \\[2ex] 
&  \qquad \qquad \qquad - \,
 \big(m + \varepsilon \phi\big) \|\dot z\|\, \delta^4(\eta- z) (\det \eta_*) \nonumber 
\end{align*}
where $\sqrt{-G} = \sqrt{-g}\,(\det \eta_*)$ and $\|\dot z\| = \sqrt{-g_{ab}(z) \dot z^a \dot z^b}$. 

Computing the SEM tensor density in this formulation, we obtain $\mathcal T^\mu{}_\nu \equiv 0$ and the other components as before. That the  spacetime components vanish is actually a consequence of the generalized Hilbert formula (3.13) in \cite{GoMa1992}, since the nondynamic fields $\eta$ and $K$ do not transform under $\textup{Diff}(X)$. (In the original formulation, the nondynamic metric $G$ on $X$ \emph{does} transform under the spacetime diffeomorphism group with the result that \eqref{sptmSEM} is nonzero.) The difference between this SEM tensor density and the previous one stems from: ($i$) the spacetime  metric no longer being regarded as a field, so that it cannot contribute to the energy, momentum, and stress content of the system, and ($ii$) the subtly different manners in which the covariance fields appear in the two formulations.

That one can encounter several SEM tensor densities for the `same' system may seem surprising, but is unavoidable and can also be regarded as different ``packaging'' of the same information. What the SEM tensor density turns out to be  depends upon what the fields are, whether they are dynamic, and precisely how they appear in the Lagrangian. And even the \emph{size} of the SEM tensor density depends upon how the system is formulated! For instance, for something as simple as a relativistic free particle, we can have a $1\times 1$ SEM tensor density (which vanishes identically)---as noted in a previous remark, or a $5 \times 5$ SEM tensor density (which does or doesn't, depending on where the spacetime metric is anchored). And in the latter case, the $5\times 5$ object reduces to the $4\times 4$ Minkowski tensor density! Thus how  the system is formulated plays a substantial role insofar as how various quantities, and in particular the SEM tensor density, are to be understood.

\section{Further Examples and Outlook} \label{ExamplesSection}

To conclude we briefly mention some other systems for which our techniques should prove helpful.  We begin by upping the dimension of the matter from 1 to 2, that is, we replace the particle by a string. For variety, we also replace the mesic interaction by an electromagnetic one.

\paragraph{Charged Strings.} We closely follow the exposition in \S2. Let $(X,G)$ and $(W = \mathbb{R}  \times B,H= - H_{\mathbb{R} } \oplus H_B)$ be 4- and 2-dimensional Lorentzian spacetimes, respectively. We consider a string  worldsheet in $X$, this being a map $z: W \to X$. We use coordinates $\big(x^\mu,\lambda^A =(\tau,\sigma)\big)$ as coordinates on $X \times W$. 
The configuration bundle $Y$ is correspondingly 
$$\Lambda^1X \times_W (X \times W)  
\times_X \textup{Lor}(X) \times_W \textup{Lor}(W) \to X \times W.$$

Assume that the string carries a charge density $\rho \colon B \to \mathbb{R}$  and interacts with a dynamic electromagnetic field described by a potential
$1$-form $A$ on $X$. We also take the metric $H$ on $W$ to be dynamic; thus we adopt the Polyakov approach as in \cite{GrScWi1987}. The action for this system is
\begin{align*}
S(A,z,H) & = -  {\displaystyle \frac14}   \int_X F^{\mu\nu}(x)F_{\mu\nu}(x) \sqrt{-G(x)}\, d^{\ps 4}\ns x \\[2ex] 
& \qquad +   \int_{X \times W} A_\mu(x)  \rho(\sigma) \frac{\partial z^{\mu}}{\partial \tau}(\tau, \sigma)\, \delta^4(x-z(\tau, \sigma))\, \sqrt{H_B(\sigma)}\, d^{\ps 4}\ns x \, d \tau \,d\sigma \\[2ex]
& \qquad -    \frac{T}{2} \int_W  H^{AB}(\lambda)\, G_{\mu\nu}(z(\lambda)) z^\mu(\lambda){}_{,A} z^\nu(\lambda){}_{,B} \sqrt{-H(\lambda)}  \,d^{\ps 2}\ns \lambda
\end{align*}
where $T$ is the tension.

As in the case of the meson, we see that $z$ takes values in $X$, which is the base for the electromagnetic field. So we need to introduce covariance fields as before. As well, we take the spacetime metric to reside on the factor of $X$ in the base. Finally, let $K$ be a nondynamic  \emph{Riemannian} metric on $W$ with total volume 1. The modified configuration bundle for the concatenated variational principle is then
$$Y \times_X X^2 \times_W \textup{Riem}(W) \to X \times W$$
and the Lagrangian reads
\begin{align*}
 \bar L(x^\mu,\tau,\sigma,A_\mu & , A_{\mu,\nu}, z^a,z^a{}_{,A},H_{AB};  \eta^a,\eta^a{}_{\nu},G^{\mu\nu}, K_{AB}) \\[2ex] 
& \quad =   - {\displaystyle \frac14}  G^{\mu\alpha} G^{\nu\beta}F_{\mu\nu}F_{\alpha\beta} \sqrt{-G}\,\sqrt{K} \\[2ex]
& \qquad \quad + A_\mu\, \kappa^\mu {}_a \rho  \,\frac{\partial z^{a}}{\partial \tau} \delta^4\big(\eta-z) (\det \eta_*)\, \sqrt{H_B} \\[2ex]
& \qquad  \quad -   \frac{T}{2} H^{AB}\, G_{\mu\nu}\, \kappa^\mu {}_a \kappa^\nu {}_b  z^a{}_{,A} \, z^b{}_{,B} \, \delta^4(\eta-z)(\det \eta_*)\sqrt{-H} .
\end{align*}

Now build the product metric $G \oplus K$ on $X \times W$. Using the Hilbert formula and then integrating as before, we compute the 6-dimensional SEM tensor density as follows: the spacetime components are
\begin{equation}
\mathfrak T^{\mu\nu} = \mathfrak T^{\mu\nu}_{\textup{EM}}  +  T\Theta^{\mu\nu}
\label{semstr}
\end{equation}
where
\begin{equation*}
\mathfrak T^{\mu\nu}_{\textup{EM}} = -\bigg(\frac14G^{\mu\nu}F_{\alpha\beta}F^{\alpha\beta} + G^{\nu\beta}F^{\alpha\mu}F_{\beta\alpha} 
\bigg)\sqrt{-G} \end{equation*}
is the free electromagnetic SEM tensor density, and 
\begin{equation*}
\Theta^{\mu\nu} =  \kappa^\mu {}_a \kappa^\nu {}_b  \, \int_W  H^{AB}\, z^a{}_{,A}\, z^b{}_{,B} \, \delta^4\big(x-\eta^{-1}(z(\sigma))\big) \, \sqrt{-H}\, d^{\ps 2}\ns \sigma
\end{equation*}
is the analogue of the Minkowski tensor density for strings.  The `extra' $\mathcal T^{\mu A}$ and $\mathcal T^{A \mu}$ components are zero and, 
after integrating over $X $, the $\mathcal T^{AB}$ subblock reduces to
\begin{equation*}
\mathfrak T^{AB} =  - {\displaystyle \frac14} \! \left( \int_X G^{\mu\alpha} G^{\nu\beta}F_{\mu\nu}F_{\alpha\beta} \sqrt{-G}\, d^{\ps 4}\ns x \right)\! K^{AB}\, \sqrt{K}.
\end{equation*}

\paragraph{Continua.} Another intriguing example that we intend to pursue in future works is a charged elastic body,  fluid, or  plasma, in which one concatenates a continuum with electromagnetism on a given background metric spacetime. Such theories will likely have significant differences with the particle and string examples presented above. Amongst these differences, we expect that, unlike   mesically or electrically charged particles, continua should have  well-defined initial value problems (see also the discussion of this point in \cite{Anderson1967}). Evidence for this can be found in works such as \cite{KuReStTe2005} and \cite{Rein1990}.

One other interesting aspect of a charged elastic body is the following. If $B$ is the body manifold, then its motion in spacetime is determined by a map $z: \mathbb{R}  \times B \to X$. The main difference from our previous two examples is that rather than the delta functions $\delta^4(\eta(x) - z(\lambda))$ we must now use characteristic functions $\chi\big(\eta^{-1}(z(\mathbb{R}  \times B))\big)$. We expect that examples such as this will be key players in the future development of the point of view given in this paper.

\section*{Appendix}

Let $M$ be a manifold with coordinates $x = (x^1,\ldots,x^m)$. Here we prove that the delta function $\delta^{\ps m}(x - x_0)$ transforms as a scalar density of weight 1.

Let $\eta: M \to M$ be a diffeomorphism and $f \in C^\infty(X)$.  On the one hand,
\begin{eqnarray*}
(f \circ \eta)(x_0) & = & \int_X (f \circ \eta)(x) \, \delta^{\ps m}(x - x_0)\,d^{\ps m}x.
\end{eqnarray*}
On the other hand, by the change of variables theorem with $y = \eta(x)$,
\begin{eqnarray*}
f(\eta(x_0)) & = & \int_X f(y) \, \delta^{\ps m}(y - \eta(x_0))\,d^{\ps m}y \\[2ex]
& = & \int_X f(\eta(x)) \, \delta^{\ps m}(\eta(x) - \eta(x_0))\,|J(x)|\,d^{\ps m}x.\end{eqnarray*}
where  $J$ is the Jacobian determinant of $\eta.$ Since $f$ is arbitrary
the desired result follows upon comparing these two formul\ae .


\bibliographystyle{marsden}

\end{document}